\documentclass[artilce,twocolumn,preprintnumbers,nofootinbib]{revtex4-2}

\usepackage{amssymb}
\usepackage{amssymb}
\usepackage{mathrsfs}
\usepackage{float}
\usepackage{bm}
\usepackage{amsmath}
\usepackage{mathdots}
\usepackage{graphicx}
\usepackage{subfigure}
\usepackage{array}
\usepackage{lipsum}
\usepackage{color}
\usepackage{epstopdf}
\usepackage{setspace}
\usepackage{datetime}
\usepackage{hyperref}

\begin{document}
	
	\title{Presence of the negative pressure in the quantum vacuum}
	
	\newcommand*{\SDUST}{Department of Physics and Institute for Theoretical Physics, \\Shandong University of Science and Technology, Qingdao 266590, Shandong,
		China}\affiliation{\SDUST}
	
	\author{Zhentao Zhang}\email{zhangzt@sdust.edu.cn}\affiliation{\SDUST}
	
\begin{abstract}We investigate the microscopic origin of the negative pressure produced by the constant energy density of the vacuum. It is shown that the zero-point photons in the quantum vacuum could generate the pressures of this type in confined spaces for the photon field. We find in particular that an anomalous radiation plays a role in the occurrence of a negative pressure from the quantum vacuum.
\end{abstract}

\maketitle
\section{Introduction}
The cosmological constant in Einstein's equations is equivalent to the existence of a constant vacuum energy density $\rho_{vac}$~\cite{Carroll}. Inevitable quantum fluctuations imply that the vacuum energy may originate from the zero-point energy of quantum fields, and this perspective is supported by the overall positive sign for the reality of the quantum vacuum~\cite{Milonni}. Aside from its profound cosmological implications, $\rho_{vac}(>0)$ has a peculiar property that it produces a negative pressure. The construction of the energy-momentum tensor of the vacuum could provide an argument for this property~\cite{Peebles}. And essentially, the negative pressure is an effect required by conservation of energy~\cite{Zee}. But these facts may not directly reveal the underlying connection between the presence of negative pressure and the quantum nature of vacuum. Nevertheless, to establish the connection in general would be a formidable task, since we do not yet understand the huge discrepancy between the observational data of $\rho_{vac}$~\cite{Planck} and the prediction of zero-point energy in the quantum vacuum. However, this does not suggest that the connection cannot be established in any circumstance; in fact, that one can obtain well-defined results from the divergent zero-point energies in the confined spaces~\cite{Casimir} might provide a hint for fulfilling such a purpose. In this work we shall show that an observable negative pressure in the quantum vacuum could occur in confined spaces for the photon field, which reveals a direct connection between the negative pressure and the quantum vacuum.  

\section{The negative pressure and the quantum vacuum}
\begin{figure}[H]
	\centering
	\includegraphics[width=7cm]{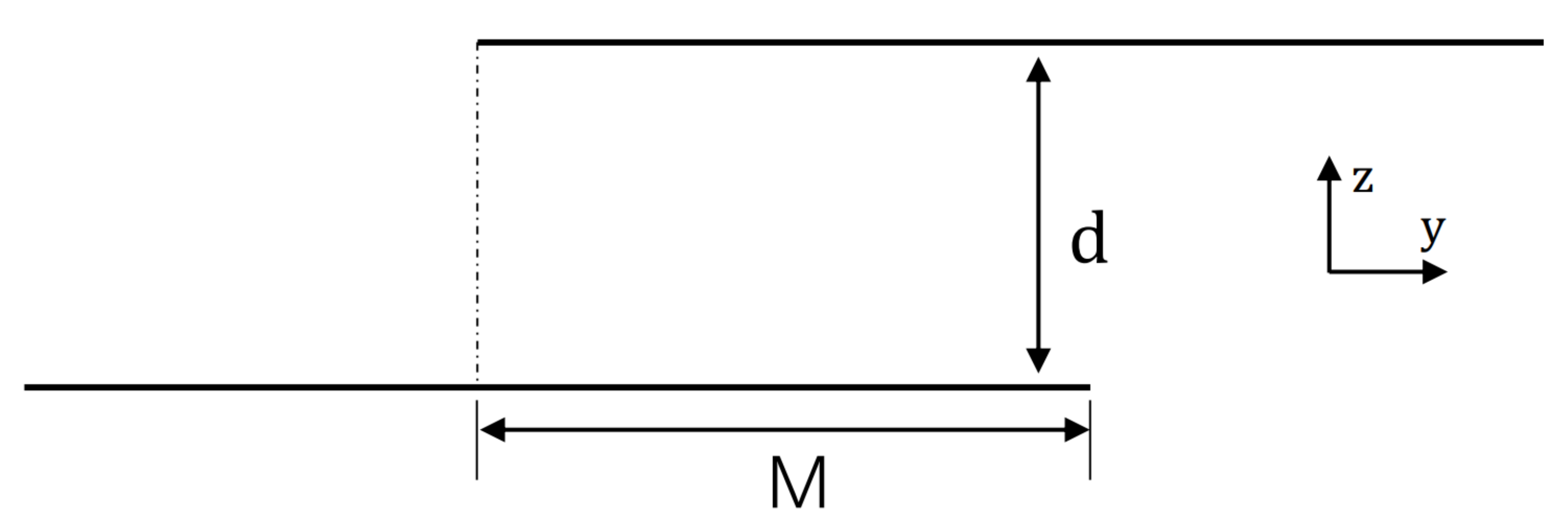}
	\caption{A side view for the parallel, perfectly conducting semi-infinite plates separated by the vacuum and overlapped partially along the $y$-direction in the Cartesian coordinate system. The overlapping distance $M$ is much larger than the separation $d$, and the infinite width of the plates along the $x$-direction will be denoted by $L$.} 
	\label{fig1}
\end{figure}

Consider the situation depicted in Fig.~\ref{fig1}. In addition to the normal Casimir force, the plates may experience an attractive tangential force between them~\cite{Zhang} 
\begin{equation}
F_T=\frac{\pi^2\hbar c}{720d^3}L,
\label{TangentialForce}
\end{equation}
along the $y$-direction, which will increase their overlapping distance. A detailed discussion on the existence of this tangential force is given in Ref.~\cite{Zhang}, which follows the conventional mechanism of the modification of zero-point energy. However, as briefly remarked by Casimir himself~\cite{Casimir} and explored in detail by Ref.~\cite{Milonni2}, there is a more straightforward interpretation of the Casimir effect: The normal Casimir force may be a consequence of the radiation pressure exerted by the virtual (i.e., zero-point) photons in the quantum vacuum, see also an investigation in a different context~\cite{Gonzalez}. As a first step to achieve the main aim of the present study, we shall show that the radiation mechanism could induce the tangential force in Eq.~(\ref{TangentialForce}), where, however, a striking anomaly will be observed.

\subsection{An anomalous vacuum radiation and the negative pressure}
The radiation interpretation comes from the fact that each virtual photon in the quantum vacuum carries a zero-point momentum $\hbar{\bf k}/2$\footnote{The zero-point momentum occurs in the quantization of the momentum operator of a field, which, like the zero-point energy, has its roots in the equal-time commutation relations of field operators \cite{Itzykson}.} and the incident field of a plane wave reflecting off a surface exerts a pressure~\cite{Milonni2}
\begin{equation} 
P=2u\cdot\text{cos}^2\theta,
\label{PositivePressure}
\end{equation}
where ${\bf k}$ is the wave vector, $u$ is the energy density of the incident field, and $\theta$ is the angle of incidence.

\begin{figure}[H]
	\centering
	\includegraphics[width=7cm]{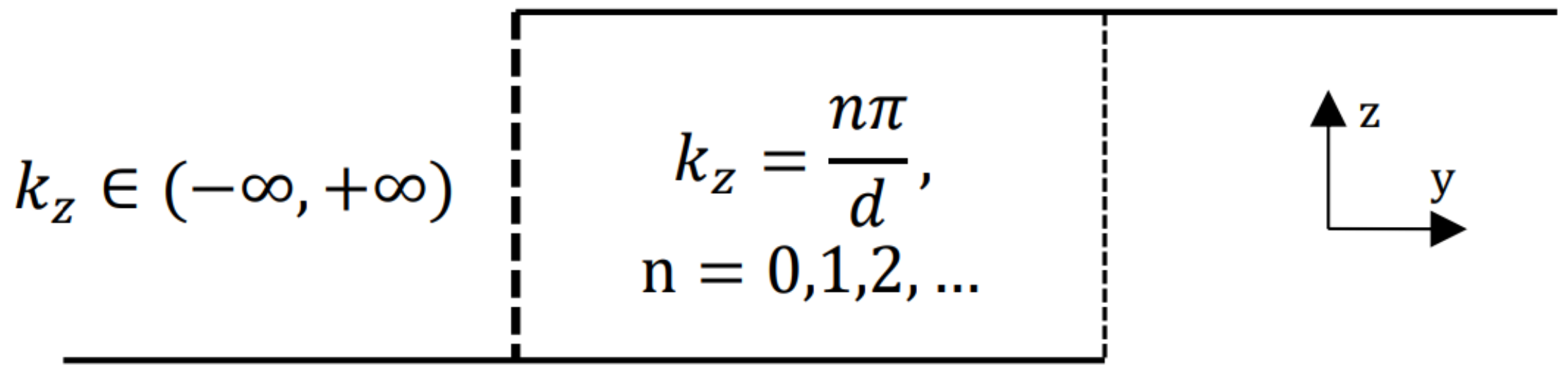}
	\caption{In the space between the conducting plates, the wavenumber $k_z$ of each mode of the field can only take on the discrete values. The modes with the other values of $k_z$ cannot pass into the space. Notice that since between the plates the $z$-components of the modes are formed by reflections off the plates, we only need to consider non-negative integer $n$ for the discrete wavenumber. Without loss of generality, in the following discussion we shall focus on the left boundary (thick dashed one).} 
	\label{fig2}
\end{figure}

The system under consideration imposes a constraint at the overlapping edges for the incident field. Only those virtual photons that carry the momenta satisfying $k_z=n\pi/d~(n=0,1,2,...)$ can pass into the space between the parallel plates, see Fig.~\ref{fig2}. The other virtual photons reflect off the edges and then a radiation pressure occurs.

Assume $u_{\omega}$ is the zero-point energy density of a mode of frequency $\omega$ in a quantization volume $V$. By
\begin{equation}
u_{\omega}= V^{-1}\frac{1}{2}\hbar\omega,~~\omega=c k,~~\text{cos}^2\theta=k_y^2/k^2
\end{equation}
for each mode, one may find a vacuum radiation pressure on the \textit{left} boundary (say) as
\begin{align}
P_{0}=\frac{\hbar c}{4\pi^3}\int_{-\infty}^{\infty}d k_x\int_{0}^{\infty}d k_y\int_{-\infty}^{\infty}d k_z\frac{k_y^2}{\sqrt{k_x^2+k_y^2+k_z^2}},
\end{align}
if all the virtual photons ($k_y>0$) reflect off the boundary, where a factor of 2 for the two independent polarizations is included. However, as explained above, the modes with the discrete values of $k_z$ would not contribute to this pressure, and we need to subtract their contribution from $P_{0}$, which reads
\begin{equation}
P'_{0}=\frac{\hbar c}{2\pi^2d}\int_{-\infty}^{\infty}d k_x\int_{0}^{\infty}d k_y\sum_{n=0}^{\infty}{'}\frac{k_y^2}{\sqrt{k_x^2+k_y^2+(n\pi/d)^2}},
\label{Transparent}
\end{equation}
where the quantization volume for $P'_{0}$ is $L_xL_yd$, and the summation $\sum{'}$ includes a $1/2$ weight factor for the $n=0$ term.

Thus, the radiation pressure is found to be
\begin{equation}
P_{vac}=P_{0}-P'_{0}.
\label{pressure}
\end{equation}

After simple algebra, the pressure may be written in the form of
\begin{align}
P_{vac}=\frac{\pi^2\hbar c}{8d^4}&\left[\int_{0}^{\infty}dn\int_{0}^{\infty}d\rho\frac{\rho}{\sqrt{\rho+n^2}}\right.\nonumber \\ &~~~~~~~\left.-\sum_{n=0}^{\infty}{'}\int_{0}^{\infty}d\rho\frac{\rho}{\sqrt{\rho+n^2}}\right].
\label{eq1}
\end{align}
Introduce a standard Casimir-type cutoff function $f(\sqrt{\rho+n^2})$ that would make each integrand in the integral decay sufficiently fast for large $\sqrt{\rho+n^2}$, and its derivatives
\begin{align}
f^{(l)}(0)=\left\{\begin{array}{c c c c}
&1  &~&l=0,\\ [10pt]
&0  &~&l=1,2,3,... \\
\end{array}\right.%
\end{align}
Then, applying the Euler-MacLaurin formula to Eq.~(\ref{eq1}), we find
\begin{equation}
P_{vac}=-\frac{\pi^2\hbar c}{720d^4}.
\label{eq3}
\end{equation}
It is seen that the result is negative, but $P_{vac}$ should be a consequence induced by the reflections of the virtual photons. Hence, we observe a striking anomaly in the sense that the virtual photons reflecting off a surface exert a negative pressure on it. Note that, to avoid ambiguity of the term ``negative pressure'' associated with a constant energy density, we will call this negative radiation pressure \textit{anomalous pressure}.

The anomalous radiation pressure should affect only the upper plate in the system since its tangential displacement causes the displacement of the left boundary. Thus, the pressure on the cross section of $Ld$ generates a force
\begin{equation}
F=P_{vac}\cdot Ld=-\frac{\pi^2\hbar c}{720d^3}L
\label{Force}
\end{equation}
acting on the upper plate, which is exactly the attractive force in Eq.~(\ref{TangentialForce}). [The minus in Eq.~(\ref{Force}) indicates that the direction of the force is opposite to the $y$-axis.] Clearly, the physics of the right boundary would be the same.

\begin{figure}[h]
	\centering
	\includegraphics[width=7cm]{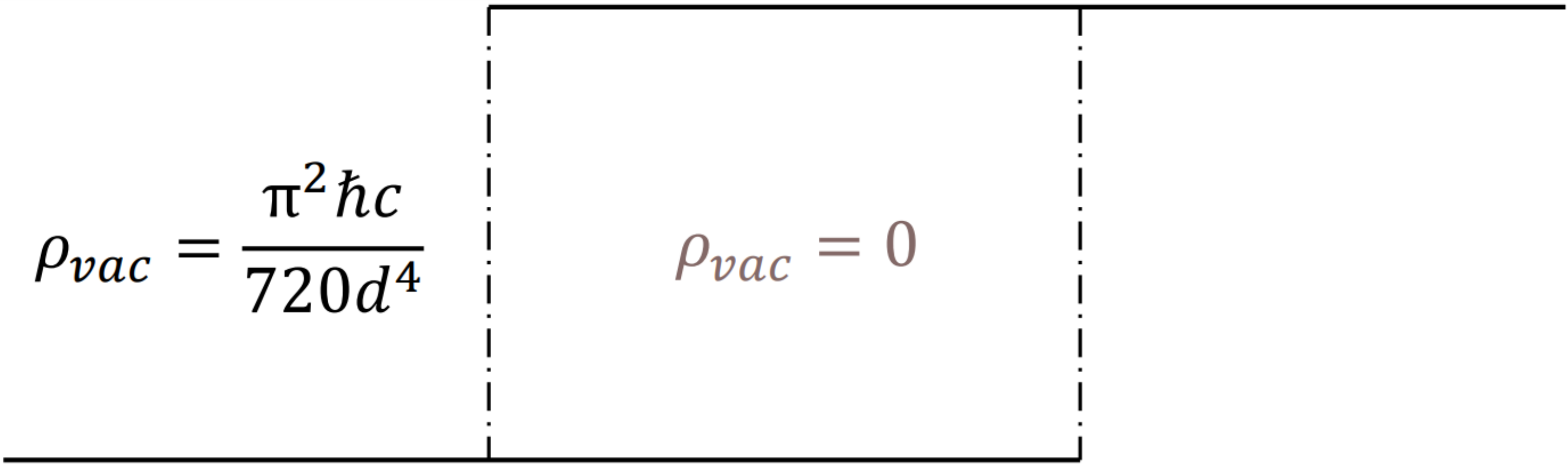}
	\caption{The expansion or contraction of a space with $\rho_{vac}=\pi^2\hbar c/(720d^4)$ caused by the tangential motion of the plates, where $\rho_{vac}$ measures the zero-point energy density difference between the free and the confined spaces for the photon field.}
	\label{fig3}
\end{figure}

However, it is crucial to realize that the misaligned system may be an unusual physical realization, in view of the fact that the tangential motion of the plates is precisely equivalent to a physical process that the expansion or contraction of a space with a finite, \textit{positive} constant vacuum energy density $\rho_{vac}=\pi^2\hbar c/(720d^4)$, see Fig.~\ref{fig3}. 

Thus, the system implies a negative pressure on the side boundary, compressing the volume with the positive constant energy density, as
\begin{equation}
P_{neg}=-\frac{d E_{vac}}{d V}=-\rho_{vac},
\label{eq4}
\end{equation}
where $E_{vac}$ is the vacuum energy of the (infinite) volume $V$, and the change of the volume is determined by the change of overlapping distance $M$. The pressure is identical to $P_{vac}$ in Eq.~(\ref{eq3}). Therefore, we may conclude that the anomalous quantum vacuum radiation could be responsible for the appearance of the negative pressure in the configuration.

\subsection{An ordinary vacuum radiation and the negative pressure}

The above discussion might lead one to speculate that the anomalous radiation may be a universal mechanism to generate the negative pressure in the quantum vacuum. However, as shown below, this may not be true. 

Consider that in the system one conducting plate is replaced with an infinitely magnetically permeable plate. This is a nontrivial theoretical construction since the normal Casimir force between the plates is now repulsive~\cite{Boyer}, which can also be regarded as an anomaly~\cite{Boyer,Hushwater}, due to the technical detail that the discrete wavenumbers in the overlapping space satisfy~\cite{Boyer}
\begin{equation}
k_z=(2n+1)\frac{\pi}{2d},~~~n=0,1,2,...
\end{equation} 
In this circumstance the radiation pressure on the left boundary should be
\begin{equation}
\tilde{P}_{vac}=P_{0}-\tilde{P}'_{0},
\label{pressure2}
\end{equation}
where
\begin{align}
\tilde{P}'_{0}=\frac{\hbar c}{2\pi^2d}&\int_{-\infty}^{\infty}d k_x\int_{0}^{\infty}d k_y\nonumber\\
&\times\sum_{n=0}^{\infty}\frac{k_y^2}{\sqrt{k_x^2+k_y^2+(n+1/2)^2\pi^2/d^2}}.
\label{Transparent2}
\end{align}
It can be cast into
\begin{align}
\tilde{P}_{vac}=\frac{\pi^2\hbar c}{8 d^4}&\left[\int_{-\frac{1}{2}}^{\infty}dn\int_{0}^{\infty}d\rho\frac{\rho f }{\sqrt{\rho+(n+\frac{1}{2})^2}}\right.\nonumber\\ &\left.~~-\sum_{n=0}^{\infty}\int_{0}^{\infty}d\rho\frac{\rho f}{\sqrt{\rho+(n+\frac{1}{2})^2}}\right],
\label{A1}
\end{align}
where $f=f(\sqrt{\rho+(n+1/2)^2})$ is a sufficiently strong cutoff function, and its derivatives
\begin{align}
f^{(l)}(1/2)=\left\{\begin{array}{c c c c}
&1  &~&l=0,\\ [10pt]
&0  &~&l=1,2,3,... \\
\end{array}\right.%
\label{A2}
\end{align}

\begin{figure}[H]
	\centering
	\includegraphics[width=7cm]{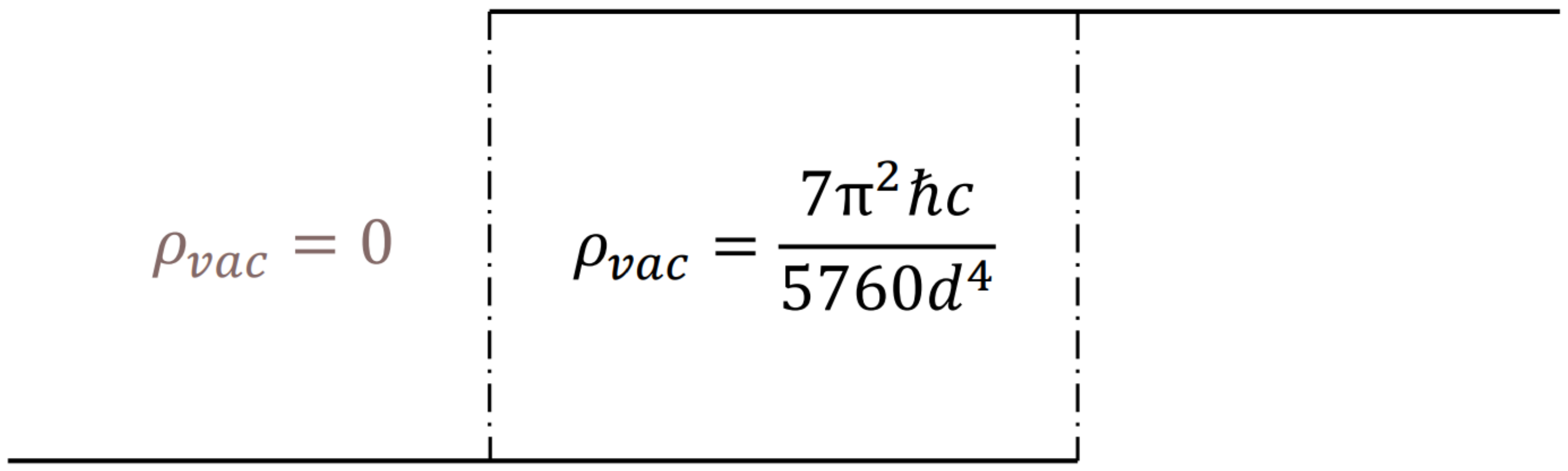}
	\caption{The expansion or contraction of a space with the vacuum energy density ${7\pi^2\hbar c}/(5760d^4)$ characterized by the tangential displacement of the plates.} 
	\label{fig4}
\end{figure}

We may obtain (see Appendix \ref{secA})
\begin{equation}
\tilde{P}_{vac}=\frac{7\pi^2\hbar c}{5760d^4},
\label{pressure21}
\end{equation}
which is a positive radiation pressure. Notice that the zero-point energy density in the space between the plates is larger than the outside, and the density difference is ${7\pi^2\hbar c}/(5760d^4)$~\cite{Boyer}.  One can see that the radiation pressure would be in agreement with the negative pressure $P_{neg}$ occurring in this misaligned system, compressing the volume, see Fig.~\ref{fig4}. It reveals that the ordinary radiation in the quantum vacuum could also generate the negative pressure required by a constant vacuum energy density.

Before we end this study, a technical note may be given here. In our discussions, the plane wave is assumed to be reflected by the single cross section at the edge, and then possible disturbance to the propagation of the incident field near the overlapping edge is not considered. Is this simplified treatment well-defined? The answer is affirmative. There are two ways to understand this conclusion: 1) From the conventional viewpoint of zero-point energy, we know that the tangential force experienced by the plates is irrelevant to the details of the quantum vacuum in the edge regions~\cite{Zhang}. This suggests that it will not be necessary to complicate our discussion on this point; 2) But a concise proof may also be outlined as follows. Notice that one can always assume an arbitrary surface outside (or inside) the space between the plates, which can form a closed surface with the cross section at the edge (including formally the ``ends'' of the infinite edge along $x$-direction, which is actually an insignificant detail), and near this imaginary surface the propagation of the virtual photons is not disturbed by the edge. Each element of the surface should experience a net, normal momentum flux of the incident field. Then, the incoming flux of momentum passing through the imaginary surface is equal to the undisturbed flux passing through the edge cross section, and it would be simple to realize that this fact should eventually approve our treatment.

\section{Summary}
In this study, we have investigated the microscopic origin of the negative pressure produced by the constant vacuum energy density. We observed that the misaligned system that consists of parallel conducting plates can provide a special physical realization that the expansion or contraction of a space with a constant vacuum energy density, and then the negative pressure should occur. We showed that the quantum vacuum radiation can be responsible for the occurrence of the negative pressures in the confined spaces for the photon field. It may be worth noting that an anomalous quantum vacuum radiation was detected in the misaligned system of conducting plates. These observations may contribute to an understanding of the negative pressure from the quantum vacuum.

\section*{Acknowledgments}
We thank R. Onofrio for the communications.

\begin{appendix}

\section{}\label{secA}
We present here a calculation of Eq.~(\ref{A1}). The equation may be rewritten as
\begin{equation}
\tilde{P}_{vac}=\frac{\pi^2\hbar c}{8 d^4}\left(-\frac{1}{2}C_1+C_2+C_3\right),
\label{A3}
\end{equation}
where
\begin{align}
C_1&=\int_{0}^{\infty}d\rho\frac{\rho f}{\sqrt{\rho+\frac{1}{4}}},  \\
C_2&=\int_{-\frac{1}{2}}^{0}d n\int_{0}^{\infty}d\rho\frac{\rho f}{\sqrt{\rho+(n+\frac{1}{2})^2}},     \\
C_3&=\int_{0}^{\infty}dn\int_{0}^{\infty}d\rho\frac{\rho f }{\sqrt{\rho+(n+\frac{1}{2})^2}} \nonumber\\&~~~~~~~~-\sum_{n=0}^{\infty}{'}\int_{0}^{\infty}d\rho\frac{\rho f}{\sqrt{\rho+(n+\frac{1}{2})^2}}.	\label{A4}
\end{align}

By the substitution $\beta=\rho+1/4$ and integration by parts,
\begin{align}
C_1&=\int_{\frac{1}{4}}^{\infty}d\beta\left(\sqrt{\beta}-\frac{1}{4\sqrt{\beta}}\right)f(\sqrt{\beta})~\nonumber \\
&=\frac{2}{3}\beta^{\frac{3}{2}}f(\sqrt{\beta})\bigg|_{\frac{1}{4}}^{\infty}-\frac{1}{2}\beta^{\frac{1}{2}}f(\sqrt{\beta})\bigg|_{\frac{1}{4}}^{\infty}\nonumber \\&~~~~~~~-\int_{\frac{1}{4}}^{\infty}\left[\frac{2}{3}\beta^{\frac{3}{2}}-\frac{1}{2}\beta^{\frac{1}{2}}\right]d f(\sqrt{\beta}).
\label{A5}
\end{align}
The third term in the second line in Eq.~(\ref{A5}) is an arbitrary constant, which will be determined by a particular choice of the cutoff function on the entire domain and thus is of no physical interest. We may get rid of this constant by adding a counterterm to the integral. After the renormalization,   
\begin{align}
C_1=\frac{1}{6}.
\label{A6}
\end{align}
Similarly, 
\begin{align}
C_2=\frac{4}{3}\int_{-\frac{1}{2}}^{0}dn (n+1/2)^3f(n+1/2)=\frac{1}{48}.
\label{A7}
\end{align}

Define
\begin{align}
F(n)=\int_{(n+\frac{1}{2})^2}^{\infty}d\alpha\left[\sqrt{\alpha}-\frac{(n+1/2)^2}{\sqrt{\alpha}}\right]f(\sqrt{\alpha})
\label{A8}
\end{align}
in Eq.~(\ref{A4}), where $\alpha=\rho+(n+1/2)^2$. Then,
\begin{align}
F'(n)=-(2n+1)\int_{(n+\frac{1}{2})^2}^{\infty}d\alpha\frac{1}{\sqrt{\alpha}}f(\sqrt{\alpha}).
\label{A9}
\end{align}
One may find
\begin{align}
F'(0)=1,\\
F^{(3)}(0)=8,
\end{align}
and $F^{(l)}(0)=0$ for $l\geqslant4$. Using the Euler-MacLaurin formula, we have
\begin{align}
C_3=\frac{1}{12}F'(0)-\frac{1}{720}F^{(3)}(0)=\frac{13}{180}.
\label{A10}
\end{align}
Therefore, the renormalized quantity
\begin{align}
-\frac{1}{2}C_1+C_2+C_3=\frac{7}{720},
\label{A11}
\end{align}
and Eq.~(\ref{pressure21}) is found.

\section{}\label{secB}
It may be of interest to note that from the expressions of virtual photon radiation in the misaligned system, a desired relation between the normal pressure $P_{vac,N}$ (see Ref.~\cite{Milonni2}) and the tangential pressure $P_{vac,T}$ can be directly obtained as
\begin{align}
P_{vac,T}=\frac{1}{2}(\rho-P_{vac,N}),
\end{align}
using a substitution 
\begin{align}
\frac{k^2_x+k^2_y}{k^2}=1-\frac{k^2_z}{k^2}
\end{align}
for each mode, where $\rho$ is the zero-point energy density difference between the continuous and the discrete wavenumbers.

\end{appendix}

\end{document}